\documentclass[floats,twocolumn,prl,preprintnumbers,shortbibliography,10pt]{revtex4-1}
\usepackage[colorlinks=true,urlcolor=blue,linkcolor=blue,citecolor=blue]{hyperref}
\usepackage{slashed,color,amsmath,amssymb,mathrsfs}
\usepackage{amsfonts,epsfig,amssymb}
\usepackage{graphicx}
\usepackage{hyperref}
\usepackage{longtable}
\usepackage{array}
\usepackage{accents}
\usepackage{tensor}
\usepackage{footmisc}
\usepackage{subfigure} 
\usepackage{autobreak}
\usepackage{longtable}
\usepackage{booktabs}
\usepackage{multirow}
\usepackage{bm}
\usepackage{subfigure}
\usepackage{chngpage}
\usepackage{ytableau}

\newcommand{\Mev}{\mathrm{MeV}}
\newcommand{\Gev}{\mathrm{GeV}}

\allowdisplaybreaks

\newcommand{\be}{\begin{equation}}
\newcommand{\ee}{\end{equation}}
\newcommand{\bea}{\begin{eqnarray}}
\newcommand{\eea}{\end{eqnarray}}
\newcommand{\sectitle}[1]{\vspace{.4cm}{\em #1.--}}
\newcommand{\sectitlequestion}[1]{\vspace{.4cm}{\em #1?--}}

\begin{document}

\title[]{New insights into the doubly charmed exotic mesons }
\author{Di Guo$^{1,2,3}$}
\author{Qin-He Yang$^{1,2}$}
\author{Ling-Yun Dai$^{1,2}$}
\email{dailingyun@hnu.edu.cn}
\author{A. P. Szczepaniak$^{4,5,6}$}
\email{aszczepa@indiana.edu}
\affiliation{$^{1}$ School of Physics and Electronics, Hunan University, Changsha 410082, China}
\affiliation{$^{2}$ Hunan Provincial Key Laboratory of High-Energy Scale Physics and Applications, Hunan University, Changsha 410082, China}
\affiliation{$^{3}$ Department of Physics, Tianjin Renai College, Tianjin 301636, China}
\affiliation{$^{4}$ Center for Exploration of Energy and Matter, Indiana University, Bloomington, IN 47403, USA}
\affiliation{$^{5}$ Physics Department, Indiana University, Bloomington, IN 47405, USA}
\affiliation{$^{6}$ Thomas Jefferson National Accelerator Facility, Newport News, VA 23606, USA}
\date{\today}

\begin{abstract}
Using effective Lagrangians constrained by the heavy quark spin symmetry and chiral symmetry, for the light quarks, we analyze the $D^0 D^0\pi^+$, $\bar{D}^0D^0\pi^0$ and $D^0\bar{D}^{*0}$ invariant mass spectra. 
Performing a simultaneous analysis of the doubly charmed and charm-anti-charm states gives further insights into the nature of the $T^+_{cc}$ and $\chi^0_{c1}(3872)$, exotic hadrons.   
It is confirmed that both states should lie below their respective $DD^*$/$D\bar{D}^*$ thresholds. Also, the contributions of the triangle and box diagrams are negligible.
\end{abstract}

\maketitle

\sectitle{Introduction}
\label{Sec:I}
Determination of hadron structure is essential for our understanding of strong interactions. 
The time-honored constituent quark model associates meson resonances with quark-antiquark bound states, but  
 this picture has difficulty explaining many of the states recently discovered in the spectrum of hadrons containing heavy quarks, {\it e.g.} the ~$\chi_{c1}(3872)/X(3872)$ \cite{CDF:2003cab}, $Z_c(3900)$ \cite{BESIII:2013ris}.
Recently another state,  referred to as $T_{cc}^+$  was found by the LHCb collaboration 
  in the invariant mass spectrum of $D^0D^0\pi^+$~\cite{LHCb:2021vvq,LHCb:2021auc}.  
Considering the valence quark components of these final states, it must contain at least four quarks {\it i.e.} at minimum 
  $|T_{cc} \rangle = |\bar{u}\bar{d}cc\rangle $.

It is possible that the $T_{cc}^+$ has connection to the $\chi_{c1}(3872)$, which was first found by Belle~\cite{Belle:2003nnu} in the $J/\psi\pi^+\pi^-$ spectrum and soon after confirmed by  CDF~\cite{CDF:2003cab}, D0~\cite{D0:2004zmu}, and BaBar~\cite{BaBar:2005xmz}. Recently, the $\chi_{c1}(3872)$ was seen directly in the invariant mass spectra of $\bar{D}^0D^0\pi^0$ by BESIII~\cite{BESIII:2020nbj} and  in $\bar{D}^0D^{*0}$ by Belle~\cite{Belle:2023zxm},  where similarity to the $T_{cc}$ is more apparent.
The $\chi_{c1}(3872)$ is also close to the open charm $D\bar D^*$ threshold,  however, the mass difference,  $\Delta M=M_{\chi_{c1}^0}-(m_{D^0}+m_{\bar{D}^{*0}})=-35\pm 60$~keV, is much smaller than that of the  $T_{cc}$ ~\cite{Hanhart:2007yq,Artoisenet:2010va,Guo:2019qcn}. 
Because these states  can, in principle, have 
both tetraquark and molecular components, they have attracted significant theoretical interest, see {\it e.g.} Refs.~\cite{Braaten:2007dw,Liu:2009qhy,Zhang:2009bv,Guo:2013zbw,Szczepaniak:2015eza,Guo:2017vcf,Guo:2019qcn,Liu:2019stu,Molina:2020kyu,Zhang:2020mpi,Wang:2017qvg,Wu:2021kbu,Dai:2021wxi,Xin:2021wcr,Wang:2023ovj,Albaladejo:2021vln,Du:2021zzh,BESIII:2023hml,Dai:2023mxm} and for some recent reviews, we refer to Refs.~\cite{Guo:2017jvc,Brambilla:2019esw,Yao:2020bxx,Chen:2022asf}.

Based on effective field theory (EFT) arguments, the heavy quarks, {\it i.e.} the two charm  quarks in the $T_{cc}$ and the charm-anti-charm pair in the $\chi_{c1}(3872)$ can be taken as  non-relativistic
\footnote{We are aware that in Refs.\cite{Shi:2020qde,Qin:2021wyh},  an application of OPE within the heavy diquark effective theory has been performed on beauty-charmed baryons $\Xi_{bc}$. }. It then permits to classify the interactions between the doubly charmed resonances and the light pseudoscalar mesons using the heavy quark spin symmetry (HQSS), $U(2N_f)$ \cite{Falk:1991nq,manohar2000heavy,Fleming:2007rp},  current algebra \cite{Gerstein:1968zz,Nutbrown:1970im,Jamin:2008rm,Dai:2019lmj}, and chiral perturbation theory \cite{Gasser:1983yg}.
This allows one to study the resonance line shape in the $DD\pi$/($D\bar{D}\pi$) and $D^0\bar{D}^{*0}$ invariant mass spectra. 
Besides, the isospin $SU(2)$ symmetry of the light quarks can be used to construct the triplet of the $T_{cc}$ and $\chi_{c1}$ states and predict the decay rates for the accompanying charge partners if they exist, $T_{cc}^{++,0}$  of the $T_{cc}^+$ and $\chi_{c1}^{\pm}$ of the $\chi_{c1}^0$ states.

\sectitle{The Effective Interactions  }
\label{Sec:II}
In the adiabatic approximation, the $cc$ pair
 in the $T^+_{cc}$ can be treated as a static, anti-triplet, spin-1 color source.
The light quark component, often referred to as \lq brown muck', i1ncludes the valence  $\bar{u}\bar{d}$ quarks and the sea of  $q\bar q$ pairs and gluons.
For both $T_{cc}^+$ and $\chi_{c1}^0$, they can be either isoscalars or isovectors. 
For the isoscalar case, the interaction Lagrangians involving the $T_{cc}^+$ are, 
\begin{eqnarray}        
\mathcal{L}_{T_{cc}}^{\rm isoscalar}\!&=&\!i g_{1}\epsilon^{ab}\epsilon_{\mu\nu\alpha\beta}v^{\alpha}T^{\beta}\langle \frac{1+v\!\!\!/}{2}\bar{\mathcal{H}}_a \gamma^{\nu} \bar{\mathcal{H}}_b^C ~\frac{1-v\!\!\!/}{2} {\gamma^{\mu}}^C \rangle  \nonumber\\
\!&+&\!g_{2} \epsilon^{ab} T^\mu \langle \Gamma_{1} \frac{1+v\!\!\!/}{2}    \bar{\mathcal{H}}_{a} \rangle \langle  \Gamma_2  \frac{1+v\!\!\!/}{2} ~\bar{\mathcal{H}}_{b}  \rangle +h.c.\,.
    \label{eq:L;THH;singlet}
\end{eqnarray}
where the subscripts $a,b$ refer to the isospin. The superscript \lq $C$' refers  to the charge conjugation, {\it i.e}   for Dirac matrices  $\Gamma^C=C^{-1}\Gamma^T C$,  with  \lq $T$'
the transpose. \lq $\langle$ $\rangle$' stands for the trace.
It is not difficult to verify that the Lagrangian is invariant under charge conjugation and parity transformation. Note that $\Gamma_{1,2}$ can be either $\gamma_{5,\mu}$ or $\gamma_{\mu,5}$, respectively.
In the effective theory with non-relativistic heavy quarks
the  two charmed, isospin  doublets ($\mathcal{H}$) are composed of the
$D$ and $D^*$ fields representing the lightest open-charm meson flavor doublet,  \cite{Casalbuoni:1996pg}:
\begin{eqnarray}
    \mathcal{H}&=&(P^*_\mu\gamma^\mu-P\gamma_5)\nonumber  \,,\\
    \bar{\mathcal{H}}&=&\gamma_0\mathcal{H}^\dagger\gamma_0=(P^{*\dagger}_\mu\gamma^\mu+P^\dagger\gamma_5)  \,,
\end{eqnarray}
with $P$, $P^*$ denoting
the  spin-$0,1$ iso-doublets  $P^{(*)} = (D^{(*),0},D^{(*),+})$ and $\bar P^{(*)} = (\bar D^{(*),0}, D^{(*),-})$, respectively, transforming as   $P^{(*)} \to P^{(*)} U^{\dag}$ and $\bar P^{(*)} \to  U \bar P^{(*)}$.
The interaction Lagrangian between the $D$ mesons and pions can be found in Ref.~\cite{Casalbuoni:1996pg}.
The Dirac algebra in the two terms in the Eq.~(\ref{eq:L;THH;singlet}) suggests  that these two Lagrangians
can be loosely associated with the coupling  
 of the  $T_{cc}$'s tetraquark and the molecular components to the open channel, respectively.
Unfortunately they both
yield identical matrix elements   between $T_{cc}$ and the $DD^*$  states thus
the individual couplings cannot be determined but only the dimensionless combination  $g_x=4\sqrt{2}(g_{1}+g_{2})f_\pi/\sqrt{m_{T_{cc}}}$. 

Similarly, one can write the effective interaction Lagrangians for the  $\chi_{c1}(3872)$ coupling with the
$D\bar{D}^*$ and $\bar{D}D^*$ states as
\begin{eqnarray}  
\mathcal{L}_{\chi_{c1}}^{\rm isoscalar}&=&i~\tilde{g}_{1}\epsilon_{\mu\nu\alpha\beta} v^{\alpha}{\chi_{c1}^{\beta}} \langle \gamma^\nu  \bar{\mathcal{H}}_a {~ \frac{1-v\!\!\!/}{2}} \gamma^\mu {~ \frac{1+v\!\!\!/}{2}} \mathcal{H}_a \rangle \nonumber\\
    &+&\tilde{g}_{2}\left(\langle  \mathcal{H}_a~ \Gamma_{1} {~ \frac{1+v\!\!\!/}{2}} \rangle \langle \Gamma_{2} \bar{\mathcal{H}}_a {~ \frac{1-v\!\!\!/}{2}}\rangle  {\chi_{c1}^{\mu}} \right. \nonumber\\
    &&-\left.\langle \Gamma_{1}~\bar{\mathcal{H}}_a  {~ \frac{1+v\!\!\!/}{2}} \rangle \langle \mathcal{H}_a \Gamma_{2}  {~ \frac{1-v\!\!\!/}{2}}\rangle  {\chi_{c1}^{\mu}}\right)\,.
    \label{eq:L;XHBH,singlet}
\end{eqnarray}
with sensitivity only to the combination
$\tilde{g}_{x}=2\sqrt{2}(\tilde{g}_{1}+\tilde{g}_{2})f_\pi/\sqrt{m_{\chi_{c1}}}$.

The $\chi_{c1}(3872)$ can also be isovector as the non-ignorable decay channels observed by experiment, ${\rm Br}[\chi_{c1}^0(\to J/\psi\rho) \to J/\psi\pi\pi]=3.8\pm1.2\%$ and ${\rm Br}[\chi_{c1}^0\to \chi_{c1}\pi^0]=3.4\pm1.6\%$ \cite{ParticleDataGroup:2022pth}.
For the isovector assumption, the $\bar{u}\bar{d}$  flavor
structure of the $T_{cc}^+$ indicates that
there may be two
charge  partners, a neutral $T^{0}_{cc}$ with the $ cc \bar u \bar u$  flavor content and a doubly charged,
$T^{++}_{cc}$ containing $cc \bar d \bar d$ quarks.
 The color triplet  \lq brown muck', in the $T_{cc}$ ground state, is expected to  have one unit of 
 total angular momentum so that  the $T_{cc}$
has spin-1 and positive parity.
The three charged vector fields form an isovector    $V^{i,\mu} $,
($i=1,2,3$)  which can be conveniently written  as a $2\times 2$ matrix
\begin{eqnarray}
    T^\mu_{ab} \!=\!  \sum_{j}  
    \frac{ [(-i\sigma_2)\sigma^j]_{ab}}{\sqrt{2}} V^{j,\mu}
    \! \equiv \!\left(\!\begin{array}{cc}
        T_{cc}^{0,\mu}                   & -\frac{T_{cc}^{+,\mu}}{\sqrt{2}} \\
        -\frac{T_{cc}^{+,\mu}}{\sqrt{2}} & T_{cc}^{++,\mu}
    \end{array}\!\right)_{ab} .
\end{eqnarray}
where the $V^{j,\mu}$'s have the isospin  composition,  $V^1=(\bar{u}\bar{u}-\bar{d}\bar{d})/\sqrt{2}$, $V^2=-i(\bar{u}\bar{u}+\bar{d}\bar{d})/\sqrt{2}$, and $V^3=(\bar{u}\bar{d}+\bar{d}\bar{u})/\sqrt{2}$, respecively.
Under flavor $SU(2)$ it   transforms as
$T^\mu \to T^\mu U^{\dag}  U^{\dag}$. 
The interaction Lagrangians describing the $T_{cc}$ coupling to the $DD^*$ are
\begin{eqnarray}    \mathcal{L}^{\rm isovector}_{T_{cc}}&=&i~g_{1}\epsilon_{\mu\nu\alpha\beta}v^{\alpha}{T^{\beta}_{ab}}\langle \frac{1+v\!\!\!/}{2} \bar{\mathcal{H}}_b \gamma^{\nu} \bar{\mathcal{H}}_a^C \frac{1-v\!\!\!/}{2} {\gamma^{\mu}}^C \rangle  \nonumber\\
    &+&g_{2} T^\mu_{ab}  \langle \Gamma_{1} \frac{1+v\!\!\!/}{2}   \bar{\mathcal{H}}_{a} \rangle \langle  \Gamma_2  \frac{1+v\!\!\!/}{2} \bar{\mathcal{H}}_{b}  \rangle +h.c.\,,    \label{eq:L;THH;Full}
\end{eqnarray}
 In this assumption, it has two charge partners,
\begin{eqnarray}
    {\chi_{c1}^\mu}_{ab} = \left(\begin{array}{cc}
            \frac{\chi_{c1}^{0,\mu}}{\sqrt{2}} & \chi_{c1}^{+,\mu}                   \\
            \chi_{c1}^{-,\mu}                  & -\frac{\chi_{c1}^{0,\mu}}{\sqrt{2}}
        \end{array}\right) \,,
\end{eqnarray}
The effective interaction of the  $\chi_{c1}$ with the
$D\bar{D}^*$ and $\bar{D}D^*$ states is given as
\begin{eqnarray}
    \mathcal{L}^{\rm isovector}_{\chi_{c1}}&=&i~\tilde{g}_{1}\epsilon_{\mu\nu\alpha\beta} v^{\mu}{\chi_{c1}^{\nu}}_{ab} \langle \gamma^\nu  \bar{\mathcal{H}}_a {~ \frac{1-v\!\!\!/}{2}} \gamma^\mu {~ \frac{1+v\!\!\!/}{2}} \mathcal{H}_b \rangle \nonumber\\
    &+&\tilde{g}_{2}\left(\langle\mathcal{H}_a~ \Gamma_{1} {~ \frac{1+v\!\!\!/}{2}} \rangle \langle \Gamma_{2} \bar{\mathcal{H}}_b {~ \frac{1-v\!\!\!/}{2}}\rangle  {\chi_{c1}^{\mu}}_{ab} \right. \nonumber\\
    &&-\left.\langle \Gamma_{1}~\bar{\mathcal{H}}_a  {~ \frac{1+v\!\!\!/}{2}} \rangle \langle \mathcal{H}_b \Gamma_{2}  {~ \frac{1-v\!\!\!/}{2}}\rangle  {\chi_{c1}^{\mu}}_{ba}\right)\,.
    \label{eq:L;XHBH;Full}
\end{eqnarray}

The interaction lagrangians of $D$, $D^*$ and light pseudoscalar meson are given in Ref.~\cite{Casalbuoni:1996pg}
\begin{eqnarray}
    \mathscr{L}_{DD}&=&i\langle H_bv^\mu D_{\mu ba}\bar{H}_a\rangle+ig\langle H_b\gamma_\mu \gamma_5\mathcal{A}^\mu_{ba}\bar{H}_a\rangle \nonumber\\ 
    &&+\frac{f_\pi^2}{4}\partial_\mu\Sigma_{ab}\partial^\mu\Sigma^{\dagger}_{ba}  \,. \label{eq:DDP}
\end{eqnarray}

\begin{figure}[htbp]
    \centering    \includegraphics[width=0.48\textwidth]{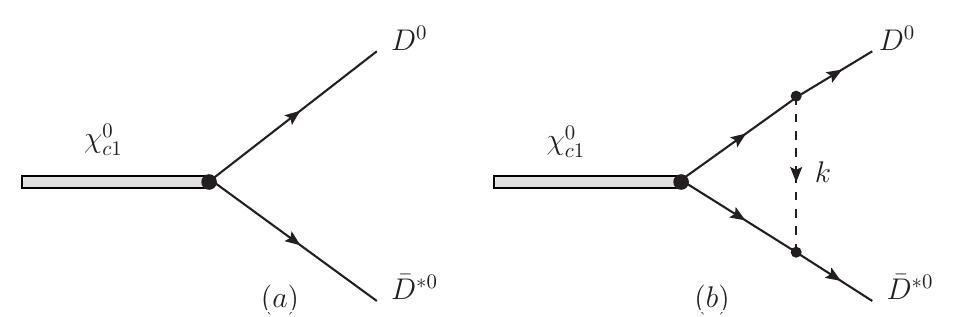}
    \caption{Feynman diagrams of $\chi_{c1}^0 \to D^0\bar{D}^{*0}$.}
    \label{Fig:diagram3872twobody}
\end{figure}

\begin{figure}[htbp]
    \centering
    \includegraphics[width=0.48\textwidth,height=0.15\textheight]{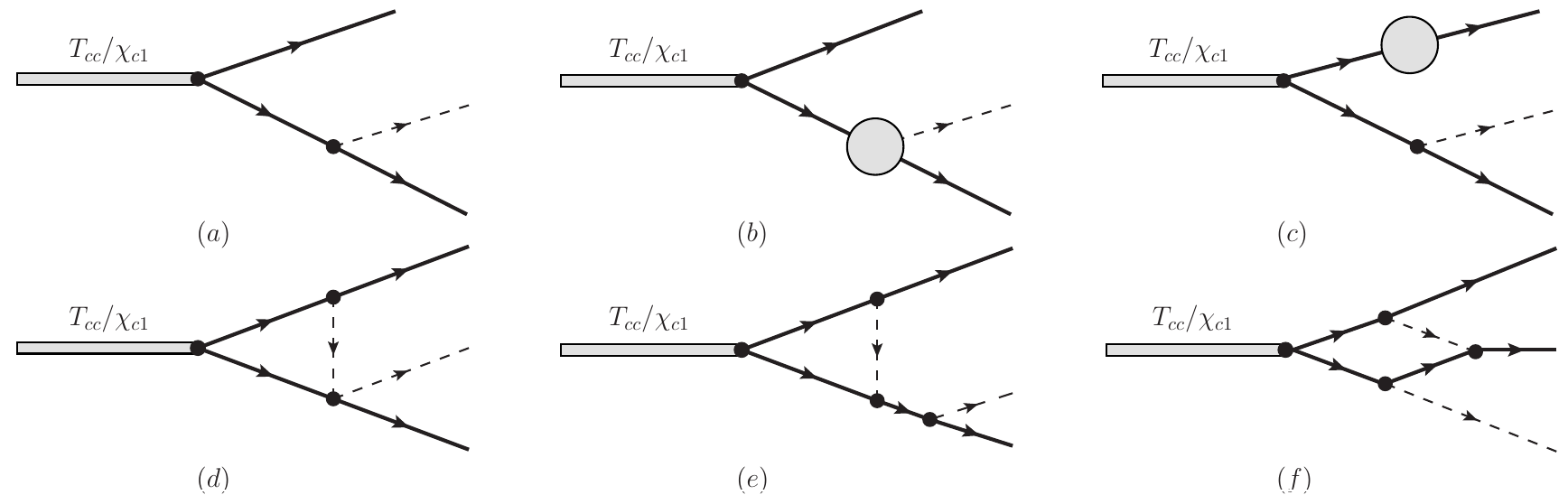}
    \caption{Feynman diagrams for the decays of $T_{cc}^{+}$ and $\chi_{c1}^0$. The solid lines represent heavy mesons, the dashed lines are the pions, and the shadow  bubble represents one-loop  loop corrections caused by a pion internal line.}
    \label{Fig:diagram}
\end{figure}

\sectitlequestion{Is $\chi_{c1}^0$ below the threshold}\label{Sec:III}
Whether the $\chi_{c1}^0$ (3872) lies below the $D^0\bar{D}^{*0}$ threshold  is a long time question \cite{Guo:2017jvc}.    
The Feynman diagrams for the two-body decay, 
$\chi_{c1}^{0}\to D^0\bar{D}^{*0}$, 
are given in Fig.~\ref{Fig:diagram3872twobody}, and the ones for the three-body decay, 
$\chi_{c1}^{0}\to D^0\bar{D}^{0}\pi^0$, are shown in Fig.~\ref{Fig:diagram}.  
The two-body decay width is given as \cite{ParticleDataGroup:2022pth}:
\begin{eqnarray}
    \Gamma(Q)=\int\frac{1}{32\pi^2}|\mathscr{M}(\chi_{c1}^0\to D^0\bar{D}^{*0})|^2\frac{|\vec{p}_1|}{Q^2}d\Omega\,, \label{Eq:DDst}
\end{eqnarray}
where $|\vec{p}_1|$ is the  breakup momentum 
  and Q is the invariant mass of $D^0\bar{D}^{*0}$. 
The three-body decay width is given in Eq.~(\ref{Eq:gammaj}) below.  

The branching ratio of $\chi_{c1}^{0}\to D^0\bar{D}^{*0}$ should be  part of the branching ratio of  $\chi_{c1}^{0}\to D^0\bar{D}^{0}\pi^0$, since $\bar{D}^{*0}$ is not an asymptotic state as it decays to $\bar{D}^0\pi^0$ by roughly two thirds. We find that the three-body decay width 
$\Gamma(\chi_{c1}^0 \to D^0\bar{D}^{0}\pi^0)  = 0.58\pm 0.23$ MeV~\cite{ParticleDataGroup:2022pth} requires $\tilde{g}_x$ to be in order one. Correspondingly, if $M_{\chi_{c1}^0}$ is above the threshold, one finds that the two-body decay width of $\Gamma(\chi_{c1}^{0}\to D^0\bar{D}^{*0})$ is too large,
specifically, contradicting the expectation that $\Gamma(\chi_{c1}^{0}\to D^0\bar{D}^{0}\pi^0)$ should be no less than $\Gamma(\chi_{c1}^{0}\to D^0\bar{D}^{*0})*{\rm Br}[\bar{D}^{*0}\to \bar{D}^0\pi^0]$.

For example,  assuming $\tilde{g}_x=2.535$ and $M_{\chi_{c1}^0}=3871.700$ (10~keV above the threshold), one finds $\Gamma(\chi_{c1}^0 \to D^0\bar{D}^{0}\pi^0)=0.777$~MeV , which is compatible with experimental value but then $\Gamma(\chi_{c1}^0 \to D^0\bar{D}^{*0})=380$~MeV, 
 which is even much larger than  total width of the $\chi_{c1}^0$, $1.19\pm 0.21$~MeV. 
Therefore, one concludes that the mass of the $\chi_{c1}^0$ should be below the $D^0\bar{D}^{*0}$ threshold to reduce the $ D^0\bar D^{*0}$  decay probability. A more careful analysis of this problem will be discussed in the next sections.

\sectitle{Reaction Amplitudes and Observables}\label{Sec:IV}
The Feynman diagrams for the amplitudes, $\mathscr{M}^{(j)}$ for   $T_{cc}^{+}$/$\chi_{c1}^0$ decaying to a specific final state $j$ that contains   $DD\pi$/$D\bar{D}\pi$ mesons are shown in Fig.~\ref{Fig:diagram}.
Note that the bottom diagrams of Fig.~\ref{Fig:diagram}, e.g. the triangle diagrams, have been partly studied in Ref.\cite{Fleming:2007rp, Dai:2023mxm} in X-EFT. 
Similar diagrams can be drawn for representative decays of the charge partners, $T_{cc}^{++} \to D^+ D^+ \pi^0$, $D^+D^0\pi^+$, $T_{cc}^{0} \to D^0 D^0 \pi^0$, $D^0D^+\pi^-$, and $\chi_{c1}^{\pm}\to D^{\pm}\bar{D}^0(D^0)\pi^0$, $D^{\pm}D^{\mp}\pi^{\pm}$, $\bar{D}^0D^0\pi^{\pm}$.
The three-body decay widths of $T_{cc}$ and $\chi_{c1}$, to the specific final states, $j$, are given by integrating over the two Dalitz-plot variables  $(s,t)$ ~\cite{ParticleDataGroup:2022pth}:
\begin{eqnarray}
    \Gamma^{(j)}(Q)=
    \int_{\gamma(Q)}  ds dt
    \frac{|\overline{\mathscr{M}^{(j)
            }}|^2}{32Q^3} \,,     \label{Eq:gammaj}
\end{eqnarray}
where $\gamma(Q)$ denotes the boundary of the Dalitz plot at fixed three-body invariant mass $Q$.
For  both $T_{cc}^{+,0,++}$ and $\chi_{c1}^{0,\pm}$, we assume that the total width is saturated
by  the sum of three-body decay widths,
\begin{eqnarray}
    \Gamma^{tot}(Q) = \sum_{j}\Gamma^{(j)}(Q) \,.
\end{eqnarray}
For $\chi_{c1}^0$, the branching ratio of $D^0\bar{D}^{0}\pi^0$ are given by PDG~\cite{ParticleDataGroup:2022pth}, so the total width is defined as $\Gamma_{\chi_{c1}^0\to D^0\bar{D}^0\pi^0}(Q)$ divided by its branching ratio,
\begin{eqnarray}
    \Gamma^{tot}_{\chi_{c1}^0}(Q)= \frac{\Gamma^{(D^0\bar{D}^0\pi^0)}(Q)}{\rm BR(\chi_{c1}^0\to D^0\bar{D}^0\pi^0)} \,.
\end{eqnarray}

As a function of the three-body mass $Q$ the experimentally  measured line shape in $Y^{(j)}(Q)$ is given by
\begin{eqnarray}
    \frac{d Y_{T_{cc}, \chi_{c1}}^{(j)}}{dQ}=N^{(j)}\!\left[\!\frac{\Gamma^{(j)}(Q)}{(Q^2-M^2)^2+[M \Gamma_{tot}(Q)]^2}\!\right]\!, \label{eq:dYdQ}
\end{eqnarray}
where $N^{(j)}$ is an arbitrary   normalization constant and
$M$ is a parameter related to the pole mass of the   $T_{cc}^{+,0,++}$ or $\chi_{c1}^{0,\pm}$ resonance. To be more specific, this is the Breit-Wigner mass of the resonance, but given that the width is narrow, it is not much different from the pole masses. 
Indeed, the pole location of $T_{cc}^{+}$ found in \cite{Dai:2021wxi}, $3874.74^{+0.11}_{-0.04} -i~0.30^{+0.05}_{-0.09}$~MeV, and the masses of the ones from Refs.\cite{Albaladejo:2021vln,Du:2021zzh,Dai:2023mxm}, are compatible with  $3874.758\pm0.055 -i~0.271\pm0.024$~MeV obtained in the present work.
Besides three-body, for  $\chi_{c1}^0$, there is also a  measurement on the invariant mass spectrum of $D^0 \bar{D}^{*0}$  \cite{Belle:2023zxm},  which we also include in the fit. 
In this case, the invariant mass spectrum is described by Eq.~(\ref{eq:dYdQ}),  but with $\Gamma^{(j)}(Q)$ replaced by $\Gamma_{\chi_{c1}^0 \to D^0\bar{D}^{*0}}(Q)$, as given by Eq.~(\ref{Eq:DDst}).

\sectitle{Results and analyses}\label{Sec:V}
For the $T_{cc}$, we fit 
our amplitudes to the $D^0D^0\pi^+$ spectrum  measured by LHCb~\cite{LHCb:2021vvq,LHCb:2021auc}, and for the  $\chi_{c1}$,  we fit the  spectrum of $D^0\bar{D}^0\pi^0$ measured by  BESIII~\cite{BESIII:2020nbj}  and that of $D^0\bar{D}^{*0}$ measured by the Belle~\cite{Belle:2023zxm}, up to 3880~MeV.
Besides the overall normalization, the remaining fit parameters are the two couplings, $g_x$ and $\tilde{g}_x$, the masses of the resonances, and the coefficients of the background in the $D^0 \bar{D}^{*0}$ invariant mass spectra of $\chi_{c1}^0$.
Consequently,  one gives predictions on the decay widths of the $T_{cc}^+$ and $\chi_{c1}^0$, as well as that of the charge partners, $T_{cc}^{++,0}$ and $\chi_{c1}^{\pm}$.

The fit results of the isoscalar case are summarized in Table \ref{tab:para}, and those of the isovector case are almost the same. See discussions below.  
The fit uncertainties were obtained using bootstrap \cite{Efron:1979bxm} {\it i.e.} by 
 averaging over multiple fits obtained by randomly sampling the experimental data within its uncertainty. 
Interestingly, the  $g_x$ and $\tilde{g}_x$ are of the same order of magnitude. 
\begin{table}[ht]
    \centering
    \begin{tabular}{ccc}
        \hline\hline
        Parameters(isoscalar)            & $T_{cc}^+$         & $\chi_{c1}^0$      \\
        \hline
        $M$     (MeV)                    & $3874.758\pm0.055$ & $3871.620\pm0.021$ \\
        $\Gamma_{tot}$  (MeV)            & $0.541\pm 0.047$   & $1.496\pm0.084$    \\
        $g_x$                            & $2.12\pm0.12$      & $\cdots$           \\
        $\tilde{g}_x$                    & $\cdots$           & $2.47\pm0.75$      \\
        $b_2$($D\bar{D}^*$)($\Gev^{-3}$) & $\cdots$           & $619175\pm145204$   \\
        $b_3$($D\bar{D}^*$)($\Gev^{-3}$) & $\cdots$           & $175960\pm73583$   \\
        $\chi^2_{\mathrm{d.o.f}}$        & 1.19               & 0.81               \\
        \hline\hline
    \end{tabular}
    \caption{Fit results correspond to $\chi^2$ per degree of freedom of 
     $1.19$ for the $T_{cc}^+$ and $0.81$ for the $\chi_{c1}^0$. 
}
    \label{tab:para}
\end{table}

\begin{figure}[ht]
    \centering
    \includegraphics[width=0.48\textwidth]{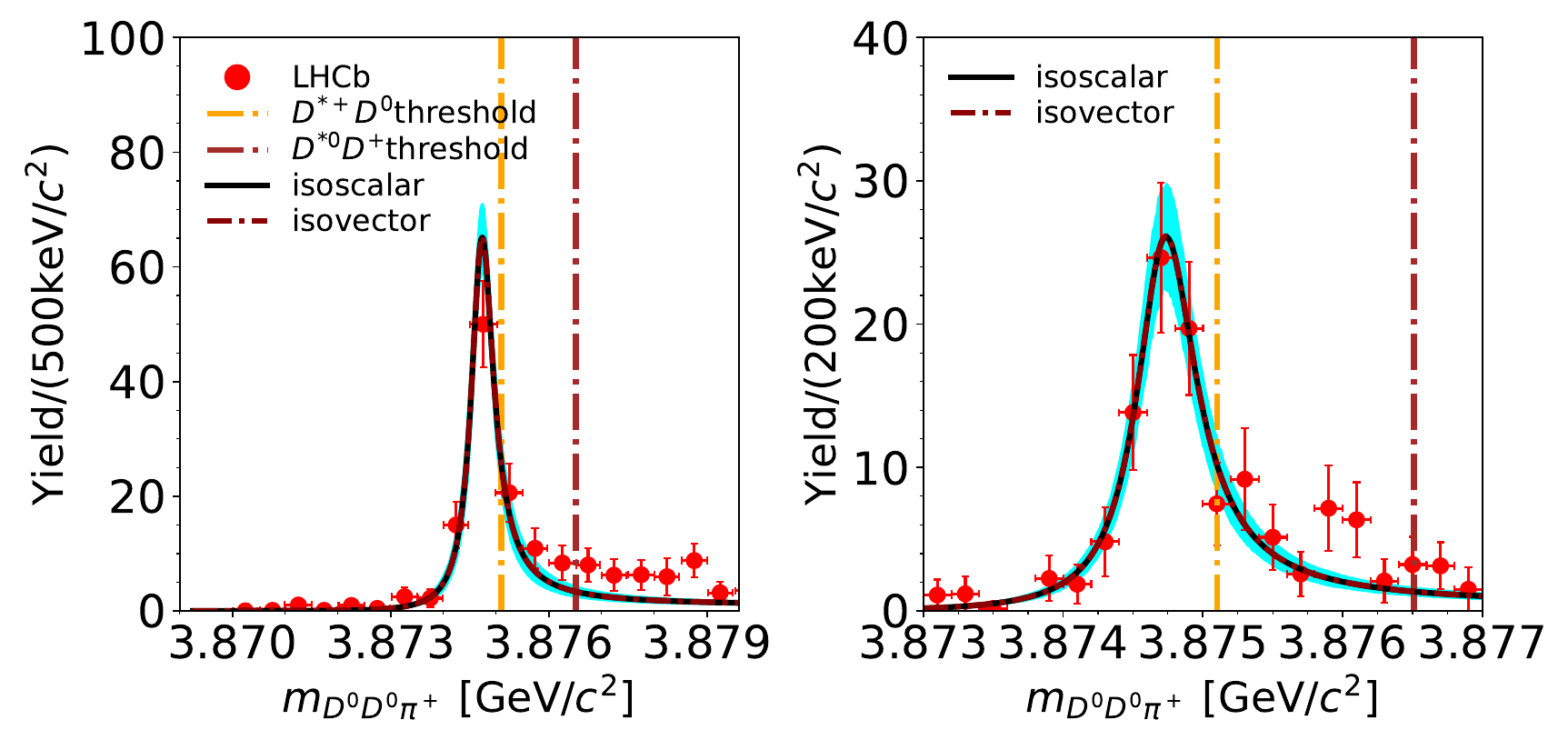}
    \caption{Our solution fits to the $D^0D^0\pi^+$ invariant mass spectra of the $T_{cc}^+$.
        The data is taken from Refs.~\cite{LHCb:2021vvq,LHCb:2021auc}.
        The cyan bands are the uncertainties taken from the bootstrap method within 2$\sigma$. \label{fig:fit}}
\end{figure}

For $T_{cc}$, the comparison of our fit to the invariant mass spectra is shown in Fig.\ref{fig:fit}.
In the right panel, the energy region around $T_{cc}^+$ is magnified. 
We find that the tree diagram, as shown in Fig.~\ref{Fig:diagram}~(a), dominates the   $T_{cc}^+\to D^0 D^0 \pi^+$ spectrum, while the triangle/box diagrams,  the diagrams at the bottom of Fig.~\ref{Fig:diagram}, are suppressed by three to four orders of magnitude. 
Especially, the interaction Lagrangian for four bosons is given as $DD\pi v\cdot\partial\pi$ in the heavy quark limit, for which the process in Fig.~\ref{Fig:diagram}~(d) vanishes to leading order. 
This finding suggests that the $T_{cc}^+$ is not a kinematical effect due to the triangle/box singularity.
Our fit gives that $M_{T_{cc}^+}=3874.758\pm0.055$~MeV and $\Gamma_{T_{cc}^+}=0.541\pm0.047$~MeV, which are close to those given by experiment \cite{LHCb:2021vvq,LHCb:2021auc}. The mass is below the $D^0D^{*+}$ threshold (3875.10~MeV), with $\Delta M=-342\pm55$ keV.

\begin{figure}[ht]
    \centering
    \includegraphics[width=0.48\textwidth]{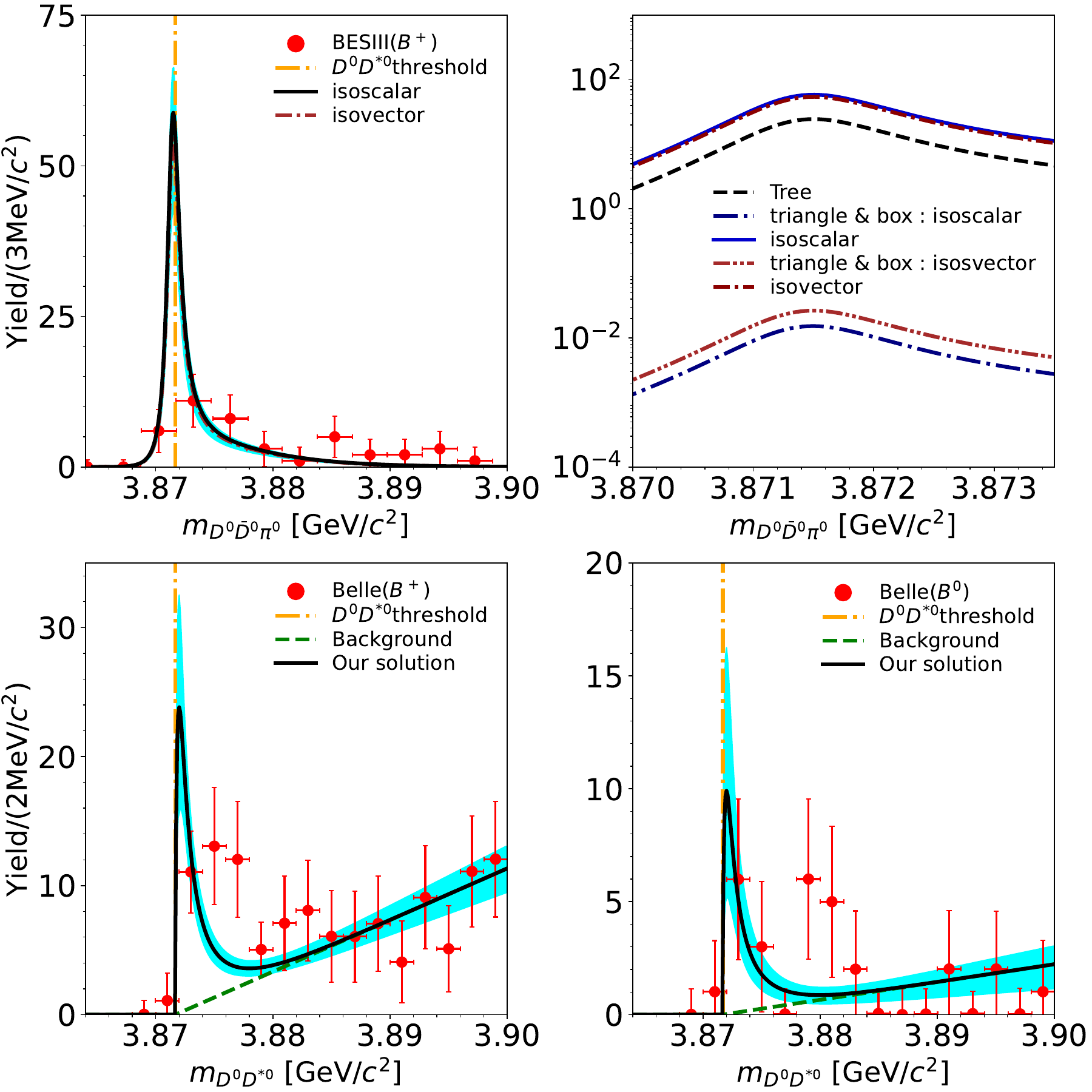}
    \caption{Our fit to the invariant mass spectra of $D^0\bar{D}^0\pi^0$ 
 from~\cite{BESIII:2020nbj,BESIII:2023hml} 
    and $D^0\bar{D}^{*0}$ from ~\cite{Belle:2023zxm}. The cyan bands correspond to a $2\sigma$ bootstrap. For the spectrum of $\chi^0_{c1}$ decaying into $D\bar{D}^*$, we add a coherent $b.g.=b(Q-m_{D^0}-m_{D^{*0}})$ to account for the rise of the data outside the signal region. 
    \label{fig:chifit} }
\end{figure}

The comparison of our results to the data for the $\chi_{c1}^0$ are shown in Fig.~\ref{fig:chifit}, with the top-left panel corresponding to $D^0\bar{D}^0\pi^0$ and the two bottom panels to the $D^0\bar{D}^{*0}$. Notice that the two-body results for isoscalars and isovectors are the same. 
In the top-right panel, we compare the total  $\bar{D}^0D^0\pi^0$  intensity 
 and the contribution from the triangle and box diagrams. 
 We  find that the  tree diagrams dominate over the triangle/box diagrams, as in the case  of the 
    $T_{cc}^+$. 
Notice that the invariant mass spectrum of $D^0\bar{D}^0\pi^0$ 
from~\cite{BESIII:2020nbj,BESIII:2023hml} is from \lq $D^{*0} \bar{D}^0$', as the $M(D\pi)$ is in the $D^{*0}$ window, and the non-resonance contribution is small.  Indeed, this is consistent with our model. The Feynman diagrams of Figs.\ref{fig:chifit} (a,b,c) indicates the contributions where the $\pi^0 D$ are from $D^{*0}$, while the other Feynman diagrams, which can be recognized as the non-resonance contribution, have much fewer contributions. See discussions below.   

The fit to the  
$D^0\bar{D}^{*0}$ spectrum from $B^+$ decays (bottom-left panel) 
  shows a discrepancy between ours and the data in the energy region of [3874-3876]~MeV. 
However, our results are still compatible with the full dataset,  also the shape is compatible with that of $D^0\bar{D}^0\pi^0$ and the $D^0\bar{D}^{*0}$ data from $B^0\to D^0\bar{D}^{*0}+K^0$. 
A combined analysis of $D^0\bar{D}^0\pi^0$ and  $D^0\bar{D}^{*0}$ data gives better constraints to refine the solution.

We have also used Eq.~(\ref{eq:dYdQ}) to estimate the invariant mass spectra of $J/\psi \pi^+\pi^-$ \cite{BESIII:2020nbj}, where the $\Gamma[{\chi_{c1}^0\to J/\psi \pi^+\pi^- }](Q)$ is calculated with an effective Lagrangian. 
It is found that our solution fits the data fairly well. This confirms the reliability of our analysis.
The decay width of $\Gamma(\chi_{c1}^0\to J/\psi \pi\pi)$ is calculated from the effective Lagrangian
$\mathscr{L}_{\chi_{c1}}=i \frac{g_J}{f_\pi^2}\epsilon^{\alpha  \beta  \mu  \nu } \psi _{\nu }\chi ^0_{\mu }(\partial _{\beta }\pi ^-\partial _{\alpha }\pi ^+-\partial _{\alpha }\pi ^-\partial _{\beta }\pi ^+)\,$. 
Notice that in the combined analysis of the experiment \cite{BESIII:2023hml}, their total width, $\Gamma_{\chi_{c1}^0}=2.67\pm1.77$~MeV is much larger than ours. That is why our line is a bit narrower than the data. Nevertheless, our results reasonably describe the $J\psi \pi \pi$ data.  These indicate that our uncertainties are reliable.

Ours gives $M_{\chi_{c1}^0}=3871.620\pm 0.021$~MeV and $\Gamma_{\chi_{c1}^0}=1.496\pm 0.084$~MeV, which are close to those of PDG
 \cite{ParticleDataGroup:2022pth}. 
Our mass does lie below the $D^0\bar{D}^{*0}$ threshold ($3871.690$~MeV), $\Delta M=-70\pm 21$~keV. This finding helps address the long-time puzzle of whether the mass of the $\chi_{c1}^0$ is below or above the $D\bar{D}^{*}$ threshold. Indeed, if we set the mass of the $\chi^0_{c1}$ above the threshold, we can not 
 simultaneously 
describe the invariant mass spectra of $D^0 \bar{D}^0\pi^0$ and $D^0\bar{D}^{*0}$.

Our width of the $\chi_{c1}^0$ reconciles the discrepancy between experiments;  the $D^0\bar{D}^{*0}$ invariant mass spectra from Belle \cite{Belle:2023zxm} gives the total decay width of $\chi_{c1}^0$ as $5.2\pm0.4\Mev$, roughly four times as larger as that of PDG \cite{ParticleDataGroup:2022pth}.
For the three-body partial decay widths, we find $\Gamma (\chi_{c1}^0\to D^0\bar{D}^0\pi^0)=0.75\pm 0.14$MeV, which is compatible with that of PDG.

Indeed, we applied the K-matrix method to fit the solutions here and extracted the poles. See Ref.~\cite{Dai:2021wxi} for formalism. The pole location is $3871.605 \pm0.013- i 0.426\pm0.004$~MeV in RS-II. It is close to that of the present analysis. 
For $T_{cc}^+$, our fit result is quite close to that of Ref.~\cite{Dai:2021wxi}, so the pole information would be similar, i.e., $3874.74^{+0.11}_{-0.04}-i0.30^{+0.05}_{-0.03}$~MeV in RS-II. 
Obviously, all the masses of these poles are still below the relevant thresholds. 

\begin{figure}[ht]
    \centering
    \includegraphics[width=0.48\textwidth]{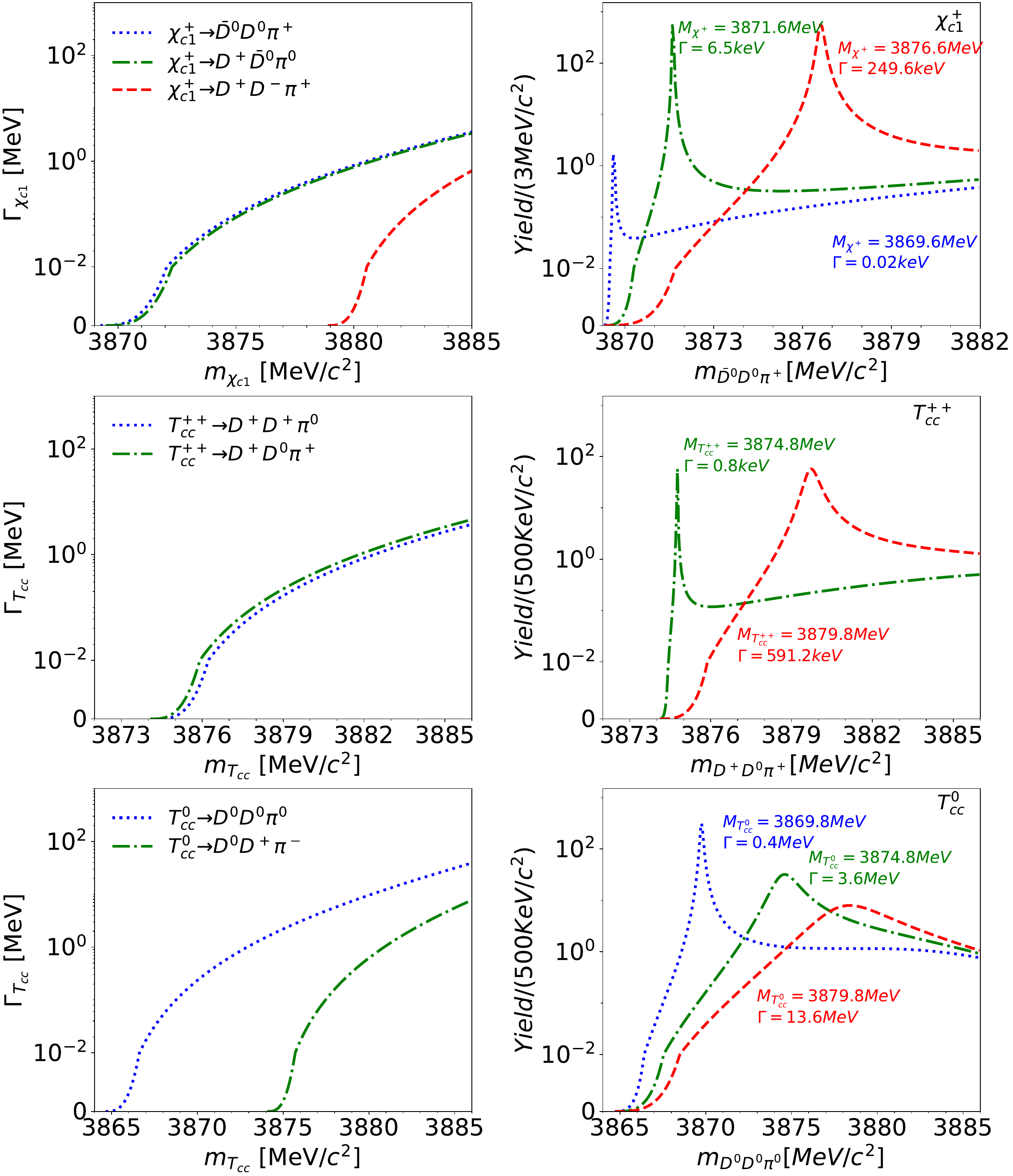}
    \caption{ The prediction of decay widths of $T_{cc}^{++,0}$ and $\chi_{c1}^{\pm}$ under the running of their masses.
        \label{Fig:gamma} }
\end{figure}

Assuming that the $T_{cc}^+$ and $\chi_{c1}^0$ are isovectors, we obtain similar results. The fit results are summarized in Table \ref{tab:para_vector}.
\begin{table}[ht]
    \centering
    \begin{tabular}{ccc}
        \hline\hline
        Parameters(isovector)            & $T_{cc}^+$         & $\chi_{c1}^0$      \\
        \hline
        $M$     (MeV)                    & $3874.758\pm0.085$ & $3871.619\pm0.034$ \\
        $\Gamma_{tot}$  (MeV)            & $0.541\pm 0.064$   & $1.496\pm0.121$    \\
        $g_x$                            & $2.85\pm0.270$      & $\cdots$           \\
        $\tilde{g}_x$                    & $\cdots$           & $2.53\pm0.276$      \\
        $b_2$($D\bar{D}^*$)($\Gev^{-3}$) & $\cdots$           & $604167\pm181372$  \\
        $b_3$($D\bar{D}^*$)($\Gev^{-3}$) & $\cdots$           & $172027\pm161120$   \\
        $\chi^2_{\mathrm{d.o.f}}$        & 1.21               & 0.81               \\
        \hline\hline
    \end{tabular}
    \caption{Fit results correspond to $\chi^2$ per degree of freedom of 
     $1.21$ for the $T_{cc}^+$ and $0.81$ for the $\chi_{c1}^0$. 
}
    \label{tab:para_vector}
\end{table}
One still has $\Delta M=-69\pm34 $~keV for $\chi_{c1}^0$ and $\Delta M=-342\pm 85$~keV for $T_{cc}^+$, and the box and triangle diagrams are found to be small compared with the tree diagrams. 
We also give predictions for 
  the decay widths of the possible charge partners of $T_{cc}^+$ and $\chi_{c1}^0$. See Fig.~\ref{Fig:gamma}.
Notice that the partners of $\chi_{c1}^0$ ($T_{cc}^+$) do not decay into $D\bar{D}$ or $D^*\bar{D}^*$ ($DD$ or $D^*D^*$) states. This is required by the conservation of parity and can be checked by expanding the Lagrangians given in Eqs.~(\ref{eq:L;THH;Full},\ref{eq:L;XHBH;Full}). 
For instance, $\chi_{c1}^0$ has a quantum number $1^{++}$, but the final states $D^0\bar{D}^0$ would have a negative parity. 
From Fig.\ref{Fig:gamma}, it can be found that the most likely decaying modes to search for the partners are  $T_{cc}^{++}\to D^0D^+\pi^+$, $T_{cc}^{0}\to D^0D^0\pi^0$, and $\chi_{c1}^{\pm}\to D^0\bar{D}^0\pi^\pm$.
We compute the invariant mass spectra of these most likely channels, as shown in the bottom of Fig.~\ref{Fig:gamma}.
The normalization factors are set to be the same as those used in the fits.  When the masses of the partners are close to the $\chi_{c1}^{0}$ or $T_{cc}^{+}$, their widths are rather small, requiring a high resolution for the experiment of LHCb, BESIII and Belle II.  For instance, when $M_{\chi_{c1}^{\pm}}=3871.6$~MeV, 
its width is only 6.5keV. 
In contrast, if the masses of these resonances are much larger than the original ones, the widths are larger, but the line shapes are more flat. One needs high statistics to find them and, if they exist, this 
 could be the reason why they have not been found yet.

\sectitle{Summary}
\label{Sec:VI}
In this paper, the effective interaction lagrangians of doubly charmed mesons and $D$/$D^*$ mesons are constructed according to heavy quark spin symmetry and $SU(2)$ chiral symmetry. 
The masses and widths of the $T_{cc}^+$ and $\chi_{c1}^0$ are extracted, $M_{T_{cc}^+}=3874.758\pm0.055$~MeV, $\Gamma_{T_{cc}^+}=0.541\pm0.047$~MeV, $M_{\chi_{c1}^0}=3871.620\pm0.021$~MeV and $\Gamma_{\chi_{c1}^0}=1.496\pm0.084$~MeV. 
Their masses do lie below the $D D^*$ or $D\bar{D}^*$ thresholds. Especially, the mass difference of the $T_{cc}^+$ is $\Delta M=M_{T_{cc}^+}-(m_{D^0}+m_{D^{*+}})=-342\pm55$~keV, and that of the $\chi_{c1}^0$ is 
\begin{equation}
\vspace{-0.2cm}
\Delta M=M_{\chi_{c1}^0}-(m_{D^0}+m_{\bar{D}^{*0}})=-70\pm21~{\rm keV} \,. \nonumber
\end{equation}
Besides, the contributions of the triangle and box diagrams are almost three to four orders smaller than those of the tree diagrams, no matter whether the $T_{cc}^+$ and $\chi_{c1}^{0}$ are isoscalars or isovectors. 
We also predict three-body decay modes of the accompanying partners if the $T_{cc}^+$ and $\chi_{c1}^0$ are isovectors, $T_{cc}^{0,++}$ and $\chi_{c1}^{\pm}$, as shown in Fig.~\ref{Fig:gamma}. These can be tested by future experiments.

\sectitle{Acknowledgements}
We thank Professors C. Meng, A. Pilloni, and S.-L. Zhang for helpful discussions. 
This work is supported by the National Natural Science Foundation of China with Grants No.12322502, 12335002, U1932110, and 11675051, Fundamental Research Funds for the central universities of China, and U.S. Department of Energy Grants No. DE-AC05-06OR23177 and No. DE-FG02- 87ER40365.

\bibliography{ref_lett}

\end{document}